\documentclass[conference]{IEEEtran}
\usepackage{cite}
\usepackage{booktabs}
\usepackage[utf8]{inputenc}
\usepackage{graphicx}
\usepackage{acronym}
\usepackage{authblk}
\usepackage{blindtext}
\usepackage{lettrine}
\usepackage{adjustbox}
\usepackage [fleqn] {amsmath}
\usepackage{cases}
\usepackage{xcolor}
\ifCLASSOPTIONcompsoc
    \usepackage[caption=false, font=normalsize, labelfont=sf, textfont=sf]{subfig}
\else
\usepackage[caption=false, font=footnotesize]{subfig}
\fi
\usepackage{tabu}
\usepackage{tikz}
\newcommand*\circled[1]{\tikz[baseline=(char.base)]{
            \node[shape=circle,draw,inner sep=0.8pt] (char) {#1};}}
\usepackage{multirow}
\usepackage{threeparttable}
\usepackage{hyperref}
\usepackage{orcidlink}

\acrodef{DL}[DL]{Deep Learning}
\acrodef{ML}[ML]{Machine Learning}
\acrodef{CDF}[CDF]{Cumulative Distribution Function}
\acrodef{DNN}[DNN]{Deep Neural Network}
\acrodef{ANN}[ANN]{Artificial Neural Network}
\acrodef{RNN}[RNN]{Recurrent Neural Network}
\acrodef{SoC}[SoC]{System on a Chip}
\acrodef{MDLS}[MDLS]{Memristive Deep Learning Systems}
\acrodef{VMM}[VMM]{Vector-Matrix Multiplication}
\acrodef{CNN}[CNN]{Convolutional Neural Network}
\acrodef{SC}[SC]{Stochastic Computing}
\acrodef{CBRAM}[CBRAM]{Conductive Bridging RAM}
\acrodef{SGD}[SGD]{Stochastic Gradient Descent}
\acrodef{ADC}[ADC]{Analog-to-Digital Converter} 
\acrodef{DAC}[DAC]{Digital-to-Analog Converter}
\acrodef{LFSR}[LFSR]{Linear Feedback Shift Register}
\acrodef{CMOS}[CMOS]{Complementary Metal Oxide Semiconductor}
\acrodef{RRAM}[RRAM]{Resistive Random Access Memory}
\acrodef{MRAM}[MRAM]{Magnetoresistive Random Access Memory}
\acrodef{IoT}[IoT]{Internet of Things}
\acrodef{MAC}[MAC]{Multiply and Accumulate}
\acrodef{RNG}[RNG]{Random Number Generator}
\acrodef{APC}[APC]{Approximate Parallel Counter}
\acrodef{IMC}[IMC]{In-Memory Computing}
\acrodef{LDMOS}[LDMOS]{Laterally-Diffused Metal Oxide Semiconductor}
\acrodef{1T1R}[1T1R]{1-Transistor-1-Resistor}
\acrodef{0T1R}[0T1R]{0-Transistor-1-Resistor}
\acrodef{BEOL}[BEOL]{Back-End-Of-The-Line}
\acrodef{DSE}[DSE]{Design Space Exploration}
\acrodef{QAT}[QAT]{Quantization-Aware Training}
\acrodef{WL}[WL]{Word Line}
\acrodef{BL}[BL]{Bit Line}
\acrodef{GPU}[GPU]{Graphics Processing Unit}
\acrodef{SPICE}[SPICE]{Simulation Program with Integrated Circuit Emphasis}
\acrodef{PVT}[PVT]{Process Voltage Temperature}

\IEEEoverridecommandlockouts
\begin{document}
\title{Design Space Exploration of Dense and Sparse Mapping Schemes for RRAM Architectures}
\author[1]{Corey Lammie\orcidlink{0000-0001-5564-1356}\thanks{\hspace{-1em}\rule{3cm}{0.5pt} \newline \textcopyright  \hspace{1pt} 2022 IEEE. Personal use of this material is permitted. Permission from IEEE must be obtained for all other uses, in any current or future media, including reprinting/republishing this material for advertising or promotional purposes, creating new collective works, for resale or redistribution to servers or lists, or reuse of any copyrighted component of this work in other works.}} % verified
\author[2]{Jason K. Eshraghian\orcidlink{0000-0002-5832-4054}} % verified
\author[3]{Chenqi Li}
\author[4]{Amirali Amirsoleimani\orcidlink{0000-0001-5760-6861}} % verified
\author[3]{Roman Genov\orcidlink{0000-0001-7506-1746}} % verified
\author[2]{Wei D. Lu\orcidlink{0000-0003-4731-1976}} % verified
\author[1]{\authorcr Mostafa Rahimi Azghadi\orcidlink{0000-0001-7975-3985}} % verified
\affil[1]{College of Science and Engineering, James Cook University, Queensland 4814, Australia}
\affil[2]{Department of Electrical Engineering and Computer Science, University of Michigan, Ann Arbor, MI 48105, USA} 
\affil[3]{Department of Electrical and Computer Engineering, University of Toronto, Toronto, Canada}
\affil[4]{Department of Electrical Engineering and Compute Science, York University, Toronto ON M3J 1P3, Canada}
\affil[$\relax$]{Email: $^1$\{corey.lammie, mostafa.rahimiazghadi\}@jcu.edu.au, $^2$jasonesh@umich.edu, $^3$chenqi.li@mail.utoronto.ca,}
\affil[$\relax$]{
$^4$amirsol@yorku.ca, $^3$roman@eecg.utoronto.ca, $^2$wluee@eecs.umich.edu}

\maketitle

\begin{abstract}
The impact of device and circuit-level effects in mixed-signal \ac{RRAM} accelerators typically manifest as performance degradation of \ac{DL} algorithms, but the degree of impact varies based on algorithmic features. These include network architecture, capacity, weight distribution, and the type of inter-layer connections. Techniques are continuously emerging to efficiently train sparse neural networks, which may have activation sparsity, quantization, and memristive noise. In this paper, we present an extended \ac{DSE} methodology to quantify the benefits and limitations of dense and sparse mapping schemes for a variety of network architectures. While sparsity of connectivity promotes less power consumption and is often optimized for extracting localized features, its performance on tiled \ac{RRAM} arrays may be more susceptible to noise due to \textit{under-parameterization}, when compared to dense mapping schemes. Moreover, we present a case study quantifying and formalizing the trade-offs of typical non-idealities introduced into \ac{1T1R} tiled memristive architectures and the size of modular crossbar tiles using the CIFAR-10 dataset. 
\end{abstract}


\IEEEpeerreviewmaketitle

\section{Introduction}
\lettrine{H}{ybrid} mixed-signal \ac{RRAM} \ac{IMC} systems are being used to efficiently perform inference of linear and unrolled convolutional layers in \ac{DL} systems~\cite{RahimiAzghadi2020,Azghadi2020,Amirsoleimani2020,Wang2020,Yao2020}. In recent years, a number of different dense and sparse mapping schemes have been proposed to reduce the required number of devices to perform inference of pre-trained \acp{ANN}~\cite{Zhou2020}. However, the efficacy of different dense and sparse mapping schemes is not well understood with respect to different circuit and device parameters, typical non-idealities, and network architectures. 

\begin{figure}[!t]
    \centering
    \includegraphics[width=0.45\textwidth]{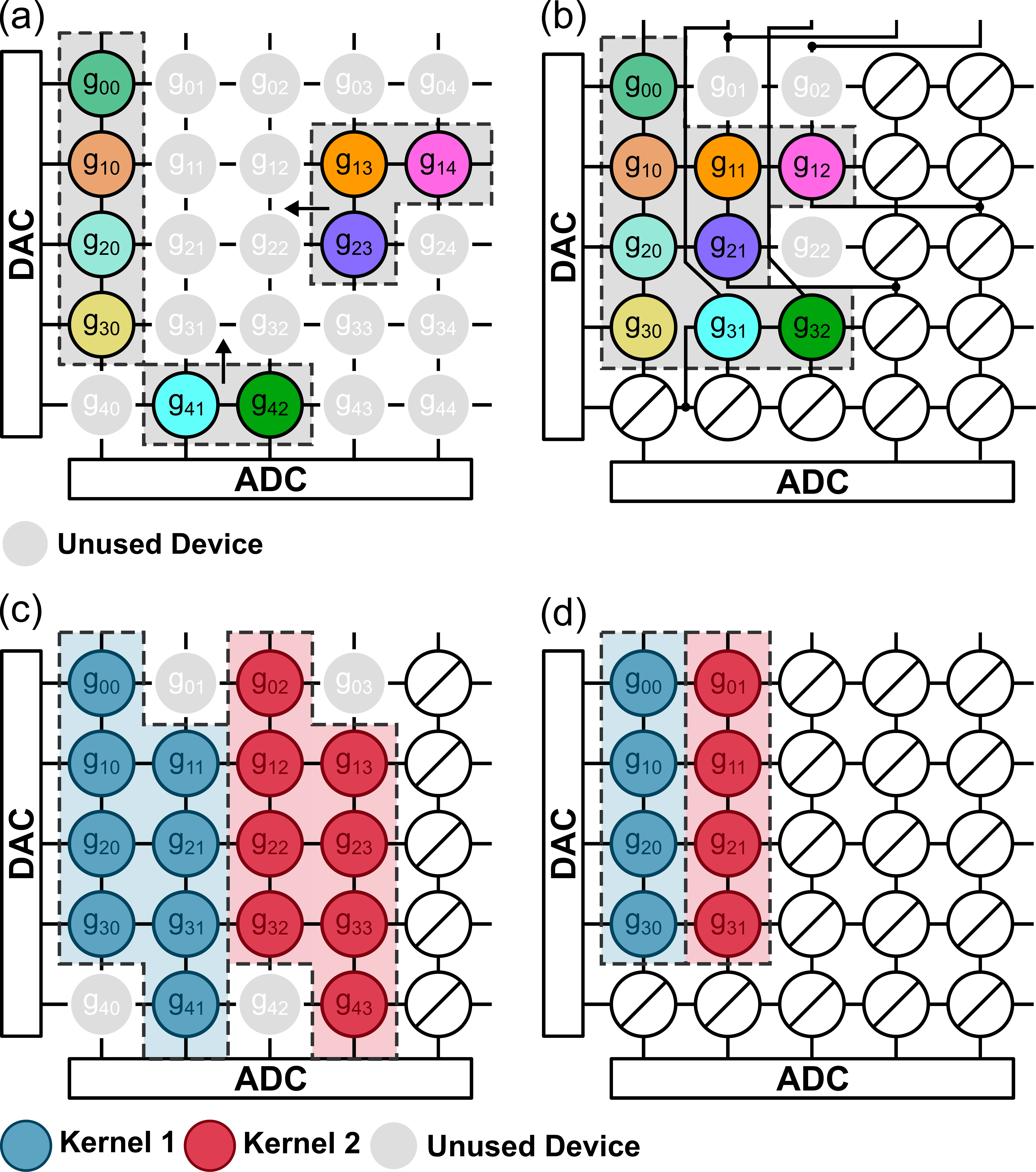}
    \caption{Overview of four popular dense and sparse mapping schemes for \ac{RRAM} architectures. When adopting a differential weight mapping scheme, (a-b) crossbar interconnects can be reconfigured to reduce the required number of devices and to reduce sparsity~\cite{Liu2021}. (c-d) For convolutional layers, kernels can either be mapped in a (c) staggered (sparse) or (d) dense arrangement, at the cost of increased read/write operations.}
    \label{fig:sparse_dense}
\end{figure}

While various hardware-aware training routines~\cite{Chen2017,Lammie2019,Fouda2020,Zhang2021} and generic search methodologies~\cite{Li2015,Krestinskaya2020} can be used to mitigate performance degradation when mapping pre-trained \ac{ANN} architectures onto \ac{RRAM} architectures, they are cumbersome, dependent on a pre-determined set of circuit and device parameters, and are not interpretable. Recently, \ac{QAT} has demonstrated the ability to improve the performance of, and to robustly reduce the error of \ac{IMC} implementations of pre-trained \acp{ANN} using \ac{RRAM} devices with discrete conductance states~\cite{Song2017,Cai2018,VanNguyen2021}.
However, it remains difficult to quantify performance trade-offs and limitations of dense and sparse mapping schemes without simulating them. In this paper, we present an extended \ac{DSE} methodology to explore the efficacy of dense and sparse mapping schemes for \ac{RRAM} architectures. Our methodology is able to explore the efficacy of dense and sparse mapping schemes for \ac{RRAM} architectures without the requirement to simulate multiple dense and sparse schemes. Using our methodology, we quantify the benefits and limitations of dense and sparse mapping schemes for four popular \ac{CNN} architectures.

\section{Related Work}
Related work in literature has attempted to quantify the various performance trade-offs in designing and implementing \ac{RRAM} architectures with respect to many different device and circuit parameters. \textit{Xu et al.}~\cite{Xu2011} studied \ac{RRAM} architecture design, and primarily focused on the choices of different peripherals to achieve the best trade-off among performance, energy, and area. \textit{Niu et al.}~\cite{Niu2012} performed a comprehensive analysis of issues related to reliability, energy consumption, area overhead, and performance. \textit{Xu et al.}~\cite{Xu2015} investigated and discussed trade-offs involving voltage drop, write latency, and data patterns. \textit{Matthew et al.}~\cite{Mathew2019} presented a \ac{DSE} framework to quantify trade-offs with respect to array sizes, write time and write energy. Recently, customizable simulation frameworks have been developed and used to simulate inference and/or training routines of \ac{RRAM} architectures~\cite{Lammie2020,Rasch2021,Peng2021,Lammie2022}.

\section{Preliminaries}
\subsection{\ac{RRAM} Crossbars}
Modular crossbar tiles, comprised of smaller sized crossbar architectures with \ac{RRAM} devices arranged in dual-column or dual-tile configurations, can be used to encode quantized analog weight-representations. By encoding $M$ \ac{WL} inputs as voltages, \acp{VMM} can be performed in $\mathcal{O}(1)$~\cite{Amirsoleimani2020}, where analog dot products are realized along each of $N$ \acp{BL}, by exploiting Ohm's law, i.e, $I_N=\sum_{i=0}^{M}V_NG_{N, M}$.

\subsection{Conventional and Sparse Mapping Schemes}
In Fig.~\ref{fig:sparse_dense}, four popular conventional sparse (a,c) and dense (b,d) weight mapping schemes are depicted. For arbitrary crossbars adopting a differential weight mapping scheme, as depicted in Fig.~\ref{fig:sparse_dense} (b), interconnects can be rerouted at the cost of increased time complexity to reduce the required number of devices~\cite{Liu2021}. Specifically when mapping convolutional layers, as depicted in Fig.~\ref{fig:sparse_dense} (d), kernels can be mapped in a dense arrangement, at the cost of increased read/write operations.

\subsection{Sparsity of Traditional \acp{ANN}}
It has been shown empirically that \acp{ANN} can tolerate high levels of sparsity, and this property has been leveraged to enable the deployment of state-of-the-art models in severely resource constrained environments, with no significant performance degradation~\cite{Gale2019, eshraghian2021training}. Sparsity is most commonly introduced using L1-regularization and Dropout layers. By increasing network sparsity, an appropriately optimized array can reduce the required number of \ac{RRAM} devices, as well as the overhead of \acp{DAC}, \acp{ADC}, and peripheral circuitry. %, is reduced.

\begin{figure}[!t]
    \centering
    \includegraphics[width=0.5\textwidth]{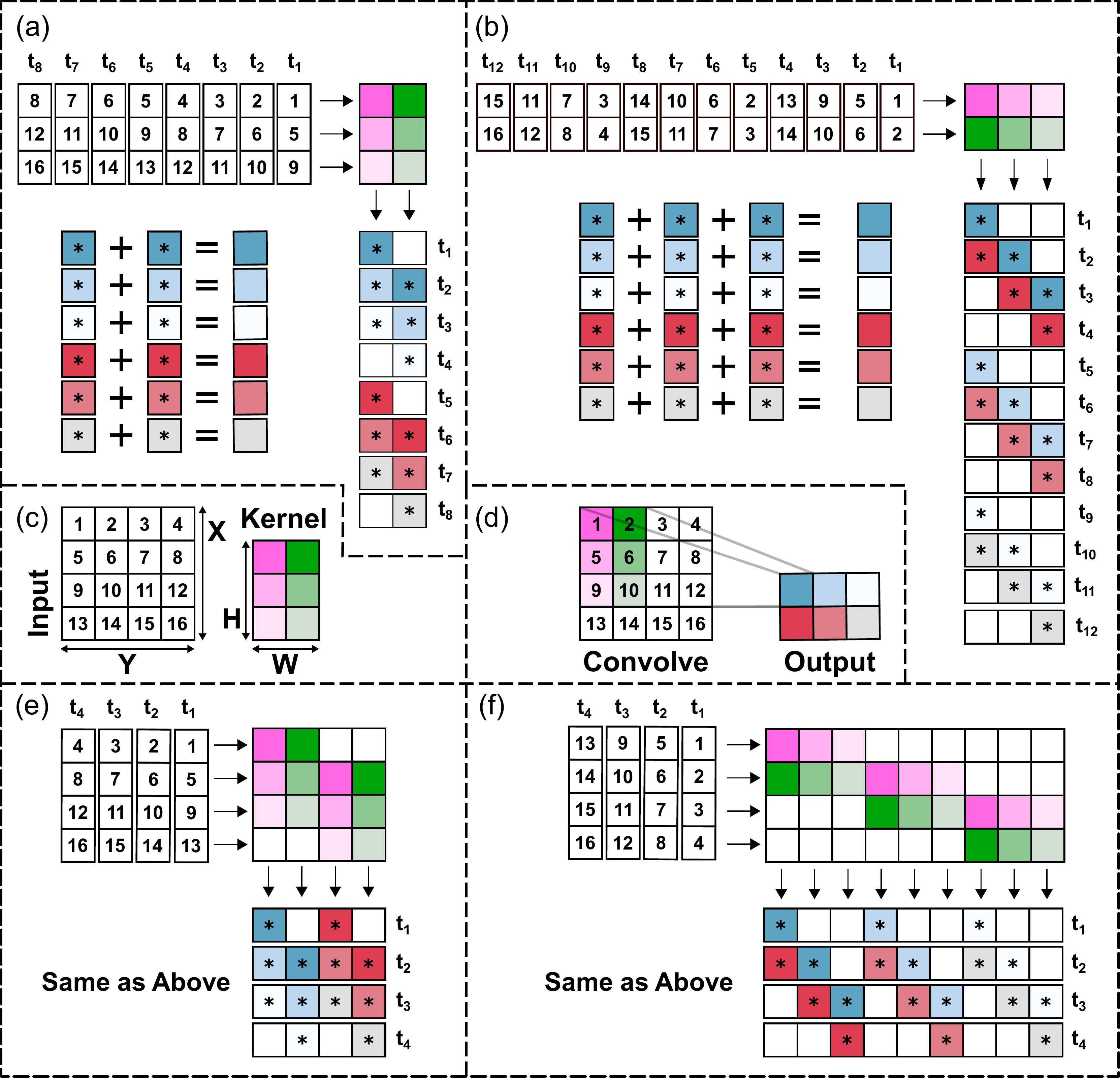}
    \caption{Dense kernel mapping computation flows for (a) vertically and (b) horizontally orientated kernels. (c-d) Labels for inputs, kernels, and outputs. Sparse (staggered) kernel mapping computation flows for (e) vertically and (f) horizontally orientated kernels.}
    \label{fig:computational_space_difference}
\end{figure}

\section{Proposed \ac{DSE} Methodology}\label{sec:proposed_DSE}
We confine our proposed \ac{DSE} search space to the following dimensions: the weight mapping scheme, I/O bit-width, tile size, maximum input encoding voltage, device/circuit non-idealities (see Section IV.B for further detail), mini-batch size, and regularization method(s). These can be categorized as network, mapping, or device/circuit related. Our proposed \ac{DSE} methodology criteria consists of the following steps:

\circled{1} For each bit-width and network architecture to investigate, \ac{QAT} is performed using a pre-determined dataset. \circled{2} Ranges of each dimension (for dimensions which are not fixed to a singular value) are determined. \circled{3} Using \ac{SPICE}-based circuit simulation tools, or \ac{RRAM}-based \ac{DL} simulation frameworks, the test or validation set accuracy for the pre-determined dataset is determined using either (a) Bayesian Optimization, or (b) a grid-search, exploring the search space. Eqs. (\ref{eq:space_requirements_sparse}) -- (\ref{eq:computational_steps_dense}) are used to determine the number of required devices, tiles, and computational steps, for each configuration, after simulating one mapping scheme. \circled{4} Contour plots are generated to explore the efficacy of different network parameters, and device/circuit parameters, using the test or validation set accuracy as the objective function. \circled{5} A score is determined for each configuration, weighting the number of required devices, tiles, and the test and/or validation set accuracy. \circled{6} Scores are manually compared.

When the scheme depicted in Fig~\ref{fig:sparse_dense}(b) was used, we assumed that convolutional kernels were mapped densely. Space requirements for convolutional and linear layers of the dense mapping schemes depicted in Fig~\ref{fig:sparse_dense}(b) can be determined using routing algorithms, where the size of modular crossbar tiles and location of zero weights are known~\cite{Liu2021}. Space requirements for convolutional layers of sparse and dense mapping schemes depicted in Fig~\ref{fig:sparse_dense}(c) and Fig~\ref{fig:sparse_dense}(d) can be determined without being physically laid out and simulated using (\ref{eq:space_requirements_sparse}) and (\ref{eq:space_requirements_dense}), respectively, where $H$, $W$, $X$, and $Y$ are defined in Fig~\ref{fig:computational_space_difference}. $D$ denotes dilation, $K$ denotes the number of kernels, and $S$ denotes the stride. For 1d-convolutional layers, $W=1$.
\begin{align}
    D_{\text{req\_sparse}} &= \frac{K^2XW[(X+2P-D(H-1)-1)]}{S+1} \label{eq:space_requirements_sparse}\\
    D_{\text{req\_dense}} &= KHW \label{eq:space_requirements_dense}
\end{align}
The required number of computational steps in Fig~\ref{fig:sparse_dense}(d) can be determined using (\ref{eq:computational_steps_dense})
\begin{equation}\label{eq:computational_steps_dense}
    C_{\text{diff\_sparse}} = \frac{X+2P-D(H-1)-1}{S+1}.
\end{equation}

\section{A Case Study}\label{sec:case_study}
In this Section, we present a case study investigating the performance of different \ac{1T1R} \ac{RRAM} architectures used to perform inference of linear and unrolled convolutional layers within popular \ac{CNN} architectures using the CIFAR-10 dataset. For all implementations, a dual-column differential weight representation scheme was adopted.

\subsection{Network Architectures and \ac{QAT}}
To ensure a sufficiently large design space was explored, we investigated the performance of four popular network architectures. A batch size of 256 and 257 training epochs were used for all implementations, with the RMSProp optimizer and a learning rate of 0.001512, which demonstrated significant performance empirically. To investigate the effect of network sparsity, for all implementations, the L1 weight-decay was set to 5e-4. The Xilinx Brevitas~\cite{Pappalardo2021} library was used in conjunction with the PyTorch~\cite{Paszke2019} \ac{ML} library to train all baseline network architectures. The weight sparsity and test set accuracy of each baseline implementation is presented in Table~\ref{table:baseline_performance}.
% , which we believe are representative of a large variety of different network configurations
% after a comprehensive exploratory search

\subsection{Device Non-Idealities}
In all simulations, the following device non-idealities were accounted for: device-to-device variability, a finite number of conductance states, and stuck $R_{\text{ON}}$ and $R_{\text{OFF}}$ devices (including those that have failed to electroform). We note that, while only three non-ideal device characteristics were investigated, more could easily be added, such as conductance drift and endurance and retention characteristics \cite{Lammie2021}.

\begin{table}[!t]
\centering
\caption{The performance and sparsity of each baseline network architecture.} 
\resizebox{0.5\textwidth}{!}{
\begin{tabular}{lrrc}
\toprule \toprule
\textbf{Architecture} & \multicolumn{1}{l}{\textbf{Bit-Width}} & \textbf{Zero Weight Values} & \textbf{Test Accuracy (\%)} \\
\midrule
\multirow{3}{*}{VGG-16~\cite{Simonyan2015}} & 4 & 189,020/15,239,872 (1.24\%) & 86.24 \\
 & 6 & 237,869/15,239,872 (1.56\%)  & 87.96 \\
 & 8 & 240,454/15,239,872 (1.58\%) & 87.88 \\ \midrule
\multirow{3}{*}{\begin{tabular}[c]{@{}l@{}}ResNeXt29\\2x64d\cite{Xie2017}\end{tabular}} & 4 & 109,100/9,103,552 (1.20\%) & 85.70 \\
 & 6 & 341,462/9,103,552 (3.75\%) & 85.89 \\
 & 8 & 190,482/9,103,552 (2.09\%) & 84.21 \\ \midrule
\multirow{3}{*}{MobileNetV2~\cite{Sandler2018}} & 4 & 74,318/2,261,824 (3.29\%) & 87.57 \\
 & 6 & 95,521/2,261,824 (4.22\%) & 88.48 \\
 & 8 & 123,501/2,261,824 (5.46\%) & 88.48 \\ \midrule
\multirow{3}{*}{GoogLeNet~\cite{Szegedy2015}} & 4 & 570,755/6,142,528 (9.29\%) & 55.19 \\
 & 6 & 102,490/6,142,528 (1.67\%) & 86.17 \\
 & 8 & 266,513/6,142,528 (4.32\%) & 86.72 \\
\bottomrule \bottomrule
\end{tabular}\label{table:baseline_performance}
}
\end{table}

\subsection{Modular Crossbar Tile Size}
For all implementations, modular symmetric crossbar tiles were used to mitigate the effects of non-ideal device and circuit characteristics. We investigated the following modular crossbar tile sizes: (32 $\times$ 32), (64 $\times$ 64), (128 $\times$ 128), and (256 $\times$ 256).

\begin{table*}[!t]
\centering
\caption{Best configuration for each unique architecture and bit-width.}
\resizebox{1\textwidth}{!}{
\begin{tabular}{lrrrrrrrrcrrr} \toprule \toprule
\multicolumn{3}{c}{\multirow{2}{*}{\textbf{Configuration}}} & \multicolumn{3}{c}{\multirow{2}{*}{\textbf{Required Devices (RD)}}} & \multicolumn{3}{c}{\multirow{2}{*}{\textbf{Read/Write Operations (RWO)}}} & \multirow{3}[5]{*}{\begin{tabular}[c]{@{}c@{}}\textbf{Test Set}\\\textbf{Accuracy (TSA)}\\\textbf{(\%)}\end{tabular}} & \multicolumn{3}{c}{\textbf{Normalized Weighted Score}} \\
\multicolumn{3}{c}{} & \multicolumn{3}{c}{} & \multicolumn{3}{c}{} &  & \multicolumn{3}{c}{\textbf{= [TSA / (RD $\times$ RWO)]'}} \\ \cmidrule{1-9}\cmidrule{11-13}
\textbf{Architecture} & \begin{tabular}[c]{@{}r@{}}\textbf{Batch Size/}\\\textbf{Tile Size}\end{tabular} & \textbf{Bit-Width} & \begin{tabular}[c]{@{}r@{}}\textbf{Sparse/}\\\textbf{Staggered}\end{tabular} & \begin{tabular}[c]{@{}r@{}}\textbf{Dense }\\\textbf{(A)}$^\diamond$\end{tabular} & \begin{tabular}[c]{@{}r@{}}\textbf{Dense }\\\textbf{(B)}$^\dagger$\end{tabular} & \begin{tabular}[c]{@{}r@{}}\textbf{\textbf{Sparse/}}\\\textbf{\textbf{Staggered}}\end{tabular} & \begin{tabular}[c]{@{}r@{}}\textbf{\textbf{Dense }}\\\textbf{\textbf{(A)}$^\diamond$}\end{tabular} & \begin{tabular}[c]{@{}r@{}}\textbf{\textbf{Dense }}\\\textbf{\textbf{(B)}$^\dagger$}\end{tabular} &  & \begin{tabular}[c]{@{}r@{}}\textbf{\textbf{Sparse/}}\\\textbf{\textbf{Staggered}}\end{tabular} & \begin{tabular}[c]{@{}r@{}}\textbf{\textbf{Dense }}\\\textbf{\textbf{(A)}$^\diamond$}\end{tabular} & \begin{tabular}[c]{@{}r@{}}\textbf{\textbf{Dense }}\\\textbf{\textbf{(B)}$^\dagger$}\end{tabular} \\ \midrule
 & 256/64 & 4 & 15,016x64x64 & 3,531x64x64 & 3,705x64x64 & 143x64 & 2,135x64 & 2,591x64 & 86.10 & \textbf{0.4947} & 0.1403 & 0.1098 \\
VGG-16 & 256/32 & 6 & 60,063x32x32 & 14,077x32x32 & 14,818x32x32 & 566x32 & 8,048x32 & 10,340x32 & 87.86 & \textbf{0.2546} & 0.0758 & 0.0557 \\
 & 256/32 & 8 & 60,063x32x32 & 14,072x32x32 & 14,818x32x32 & 566x32 & 8,121x32 & 10,340x32 & 87.74 & \textbf{0.2543} & 0.0748 & 0.0557 \\ \midrule
 & 256/64 & 4 & 243,653x64x64 & 2,184x64x64 & 2,213x64x64 & 657x64 & 24,266x64 & 24,712x64 & 83.87 & 0.0055 & \textbf{0.0186} & 0.0180 \\
ResNeXt29 & 64/64 & 6 & 243,653x64x64 & 2,070x64x64 & 2,213x64x64 & 657x64 & 22,681x64 & 24,712x64 & 86.08 & 0.0057 & \textbf{0.0217} & 0.0185 \\
 & 256/64 & 8 & 243,653x64x64 & 2,165x64x64 & 2,213x64x64 & 657x64 & 24,712x64 & 24,712x64 & 82.28 & 0.0054 & \textbf{0.0180} & 0.0176 \\ \midrule
 & 256/64 & 4 & 48,590x64x64 & 528x64x64 & 548x64x64 & 220x64 & 2,181x64 & 2,406x64 & 76.85 & 0.0878 & \textbf{0.8251} & 0.7182 \\
MobileNetV2 & 256/32 & 6 & 194,257x32x32 & 2,092x32x32 & 2,191x32x32 & 875x32 & 8,525x32 & 9,524x32 & 88.71 & 0.0505 & \textbf{0.4914} & 0.4201 \\
 & 256/64 & 8 & 48,590x64x64 & 516x64x64 & 548x64x64 & 220x64 & 2,119x64 & 2,406x64 & 88.52 & 0.1014 & \textbf{1.0000} & 0.8283 \\ \midrule
 & 16/64 & 4 & 561,915x64x64 & 1,345x64x64 & 1,481x64x64 & 1,246x64 & 15,142x64 & 17,552x64 & 55.54 & 0.0000 & \textbf{0.0327} & 0.0254 \\
GoogLeNet & 16/256 & 6 & 35,120x256x256 & 91x256x256 & 92x256x256 & 79x256 & 1,068x256 & 1,127x256 & 85.58 & 0.0050 & \textbf{0.1691} & 0.1584 \\
 & 16/128 & 8 & 140,479x128x128 & 354x128x128 & 371x128x218 & 312x128 & 4,207x128 & 4,417x128 & 86.58 & 0.0021 & \textbf{0.0888} & 0.0470 \\ \bottomrule \bottomrule
\end{tabular}
}\label{table:best_configurations}
\begin{tablenotes}
      \item {$^\diamond$Dense (A) refers to the dense mapping scheme depicted in Fig.~\ref{fig:sparse_dense} (b). $^\dagger$Dense (B) refers to the dense mapping scheme depicted in Fig.~\ref{fig:sparse_dense} (d). 'Min-max normalization of weight scores is performed to aid comparisons.}
\end{tablenotes}
\end{table*}

\begin{figure*}[!t]
    \centering
    \includegraphics[width=1\textwidth]{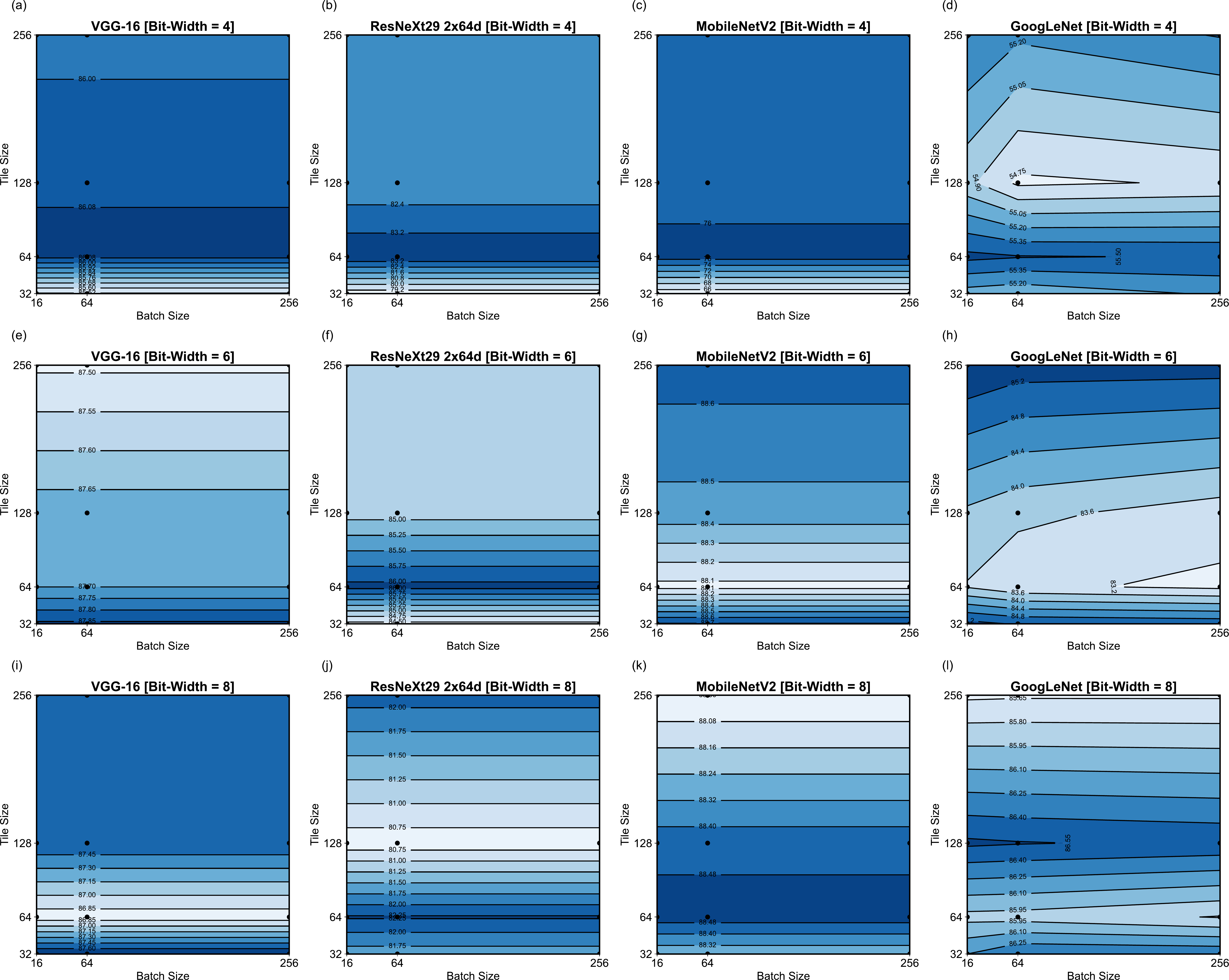}
    \caption{Contour plots depicting the dependency of the test set accuracy on symmetrical tile sizes, the batch size, and the bit-width of simulated \ac{RRAM} architectures, for each network architecture. Different tile shapes and batch sizes are explored for bit-widths of (a-d) 4, (e-h) 6, and (i-l) 8, respectively. Fixed seed values of 0 were used to ensure the same stochastic non-idealities were sampled during each simulation.}
    \label{fig:contour_plots}
\end{figure*}

\subsection{Results}
To apply our \ac{DSE} methodology, we determined the optimal batch size and tile shape (of symmetric tiles) for inference, as depicted in Fig.~\ref{fig:contour_plots}. In addition, we reported the best of a normalized device-read/write-accuracy weighted score, alongside: the required number of devices, required number of read/write operations, and the test set accuracy. For each architecture and bit-width, the optimal batch and tile sizes were determined with respect to the test set accuracy. The performance of these configurations are presented in Table~\ref{table:best_configurations}.
To simulate \ac{RRAM} architectures, the MemTorch~\cite{Lammie2020} simulation framework was used. The following fixed parameters were used to reduce the dimensionality of the explored design space: $R_{\text{OFF}}$ was sampled from a normal distribution with a mean of 100k$\Omega$ and a standard deviation of 10,000. $R_{\text{ON}}$ was sampled from a normal distribution with a mean of 10k$\Omega$ and a standard deviation of 1,000. A maximum encoding input voltage of 0.3V was used~\cite{Shim2020}, with a failure rate of stuck devices to $R_{\text{ON}}$ of 0.5\%, and a failure rate of stuck devices to $R_{\text{OFF}}$ of 0.5\%. 
The range of terms used to compute the weighted score can be standardized to reduce biases, and depending on specific user requirements, different weightings can be applied.

\section{Discussion and Conclusions}\label{sec:discussion}
In Table~\ref{table:best_configurations}, it can be observed that MobileNetV2 with a batch size of 256 and a tile size of 64 achieved the best normalized weighted score. As can be seen in Fig.~\ref{fig:contour_plots}, for all network architectures other than GoogLeNet, the batch size used during inference had a negligible influence on the test set accuracy. Our empirical results indicate that for the investigated networks, symmetrical tiles with a size of $\geq64$ where deemed optimal. For all network architectures, the optimal tile shape was found to be dependent on both the network architecture used and the bit-width. This suggests that the optimal tile size for pre-trained \acp{CNN} cannot easily be determined without performing an exploratory analysis. Despite being limited in scope, we believe that the case study in Section~\ref{sec:case_study} demonstrates the effectiveness of our presented \ac{DSE} methodology to investigate and determine dependencies between different search space dimensions.

\bibliographystyle{IEEEtran}
\bibliography{References}

\begin{thebibliography}{10}
\providecommand{\url}[1]{#1}
\csname url@samestyle\endcsname
\providecommand{\newblock}{\relax}
\providecommand{\bibinfo}[2]{#2}
\providecommand{\BIBentrySTDinterwordspacing}{\spaceskip=0pt\relax}
\providecommand{\BIBentryALTinterwordstretchfactor}{4}
\providecommand{\BIBentryALTinterwordspacing}{\spaceskip=\fontdimen2\font plus
\BIBentryALTinterwordstretchfactor\fontdimen3\font minus
  \fontdimen4\font\relax}
\providecommand{\BIBforeignlanguage}[2]{{%
\expandafter\ifx\csname l@#1\endcsname\relax
\typeout{** WARNING: IEEEtran.bst: No hyphenation pattern has been}%
\typeout{** loaded for the language `#1'. Using the pattern for}%
\typeout{** the default language instead.}%
\else
\language=\csname l@#1\endcsname
\fi
#2}}
\providecommand{\BIBdecl}{\relax}
\BIBdecl

\bibitem{RahimiAzghadi2020}
M.~Rahimi~Azghadi, Y.-C. Chen, J.~K. Eshraghian, J.~Chen, C.-Y. Lin,
  A.~Amirsoleimani, A.~Mehonic, A.~J. Kenyon, B.~Fowler, J.~C. Lee, and Y.-F.
  Chang, ``{Complementary Metal-Oxide Semiconductor and Memristive Hardware for
  Neuromorphic Computing},'' \emph{Advanced Intelligent Systems}, vol.~2,
  no.~5, p. 1900189, 2020.

\bibitem{Azghadi2020}
M.~R. Azghadi, C.~Lammie, J.~K. Eshraghian, M.~Payvand, E.~Donati,
  B.~Linares-Barranco, and G.~Indiveri, ``{Hardware Implementation of Deep
  Network Accelerators Towards Healthcare and Biomedical Applications},''
  \emph{IEEE Transactions on Biomedical Circuits and Systems}, vol.~14, no.~6,
  pp. 1138--1159, 2020.

\bibitem{Amirsoleimani2020}
A.~Amirsoleimani, F.~Alibart, V.~Yon, J.~Xu, M.~R. Pazhouhandeh, S.~Ecoffey,
  Y.~Beilliard, R.~Genov, and D.~Drouin, ``{In-Memory Vector-Matrix
  Multiplication in Monolithic Complementary
  Metal–Oxide–Semiconductor-Memristor Integrated Circuits: Design Choices,
  Challenges, and Perspectives},'' \emph{Advanced Intelligent Systems}, vol.~2,
  no.~11, p. 2000115, 2020.

\bibitem{Wang2020}
X.~Wang, M.~A. Zidan, and W.~D. Lu, ``{A Crossbar-Based In-Memory Computing
  Architecture},'' \emph{IEEE Transactions on Circuits and Systems I: Regular
  Papers}, vol.~67, no.~12, pp. 4224--4232, 2020.

\bibitem{Yao2020}
P.~Yao, H.~Wu, B.~Gao, J.~Tang, Q.~Zhang, W.~Zhang, J.~J. Yang, and H.~Qian,
  ``{Fully Hardware-Implemented Memristor Convolutional Neural Network},''
  \emph{Nature}, vol. 577, no. 7792, pp. 641--646, Jan. 2020.

\bibitem{Zhou2020}
Y.~Zhou, B.~Gao, C.~Dou, M.-F. Chang, and H.~Wu, ``{Chapter 14 - RRAM-Based
  Coprocessors for Deep Learning},'' in \emph{Memristive Devices for
  Brain-Inspired Computing}, ser. Woodhead Publishing Series in Electronic and
  Optical Materials, S.~Spiga, A.~Sebastian, D.~Querlioz, and B.~Rajendran,
  Eds.\hskip 1em plus 0.5em minus 0.4em\relax Woodhead Publishing, 2020, pp.
  363--395.

\bibitem{Liu2021}
\BIBentryALTinterwordspacing
F.~Liu, W.~Zhao, Y.~Zhao, Z.~Wang, T.~Yang, Z.~He, N.~Jing, X.~Liang, and
  L.~Jiang, ``{SME:} {ReRAM-based Sparse-Multiplication-Engine to Squeeze-Out
  Bit Sparsity of Neural Network},'' \emph{CoRR}, vol. abs/2103.01705, 2021.
  [Online]. Available: \url{https://arxiv.org/abs/2103.01705}
\BIBentrySTDinterwordspacing

\bibitem{Chen2017}
L.~Chen, J.~Li, Y.~Chen, Q.~Deng, J.~Shen, X.~Liang, and L.~Jiang,
  ``{Accelerator-Friendly Neural-Network Training: Learning Variations and
  Defects in RRAM Crossbar},'' in \emph{Proceedings of the Design, Automation
  Test in Europe Conference Exhibition (DATE)}, 2017, pp. 19--24.

\bibitem{Lammie2019}
C.~Lammie, O.~Krestinskaya, A.~James, and M.~R. Azghadi, ``{Variation-aware
  Binarized Memristive Networks},'' in \emph{Proceedings of the IEEE
  International Conference on Electronics, Circuits and Systems (ICECS)}, 2019,
  pp. 490--493.

\bibitem{Fouda2020}
M.~E. Fouda, S.~Lee, J.~Lee, G.~H. Kim, F.~Kurdahi, and A.~M. Eltawi, ``{IR-QNN
  Framework: An IR Drop-Aware Offline Training of Quantized Crossbar Arrays},''
  \emph{IEEE Access}, vol.~8, pp. 228\,392--228\,408, 2020.

\bibitem{Zhang2021}
B.~Zhang, L.-Y. Chen, and N.~Verma, ``{Neural Network Training With Stochastic
  Hardware Models and Software Abstractions},'' \emph{IEEE Transactions on
  Circuits and Systems I: Regular Papers}, vol.~68, no.~4, pp. 1532--1542,
  2021.

\bibitem{Li2015}
B.~Li, L.~Xia, P.~Gu, Y.~Wang, and H.~Yang, ``{Merging the Interface: Power,
  Area and Accuracy Co-Optimization for RRAM Crossbar-Based Mixed-Signal
  Computing System},'' in \emph{Proceedings of the Design, Automation Test in
  Europe Conference Exhibition (DATE)}, ser. DAC '15.\hskip 1em plus 0.5em
  minus 0.4em\relax New York, NY, USA: Association for Computing Machinery,
  2015.

\bibitem{Krestinskaya2020}
O.~Krestinskaya, K.~N. Salama, and A.~P. James, ``{Automating Analogue AI Chip
  Design with Genetic Search},'' \emph{Advanced Intelligent Systems}, vol.~2,
  no.~8, p. 2000075, 2020.

\bibitem{Song2017}
C.~Song, B.~Liu, W.~Wen, H.~Li, and Y.~Chen, ``{A Quantization-Aware
  Regularized Learning Method in Multilevel Memristor-Based Neuromorphic
  Computing System},'' in \emph{Proceedings of the IEEE Non-Volatile Memory
  Systems and Applications Symposium (NVMSA)}, 2017.

\bibitem{Cai2018}
Y.~Cai, T.~Tang, L.~Xia, M.~Cheng, Z.~Zhu, Y.~Wang, and H.~Yang, ``{Training
  Low Bitwidth Convolutional Neural Network on RRAM},'' in \emph{Proceedings of
  the Asia and South Pacific Design Automation Conference (ASP-DAC)}, 2018, pp.
  117--122.

\bibitem{VanNguyen2021}
T.~Van~Nguyen, J.~An, and K.-S. Min, ``{Comparative Study on Quantization-Aware
  Training of Memristor Crossbars for Reducing Inference Power of Neural
  Networks at The Edge},'' in \emph{Proceedings of the International Joint
  Conference on Neural Networks (IJCNN)}, 2021.

\bibitem{Xu2011}
C.~Xu, X.~Dong, N.~P. Jouppi, and Y.~Xie, ``{Design Implications of
  Memristor-Based RRAM Cross-Point Structures},'' in \emph{Proceedings of the
  Design, Automation Test in Europe Conference Exhibition (DATE)}, 2011.

\bibitem{Niu2012}
D.~Niu, C.~Xu, N.~Muralimanohar, N.~P. Jouppi, and Y.~Xie, ``{Design Trade-Offs
  for High Density Cross-Point Resistive Memory},'' in \emph{Proceedings of the
  ACM/IEEE International Symposium on Low Power Electronics and Design}, New
  York, NY, USA, 2012, p. 209–214.

\bibitem{Xu2015}
C.~Xu, D.~Niu, N.~Muralimanohar, R.~Balasubramonian, T.~Zhang, S.~Yu, and
  Y.~Xie, ``{Overcoming the Challenges of Crossbar Resistive Memory
  Architectures},'' in \emph{Proceedings of the IEEE International Symposium on
  High Performance Computer Architecture (HPCA)}, 2015, pp. 476--488.

\bibitem{Mathew2019}
D.~M. Mathew, A.~L. Chinazzo, C.~Weis, M.~Jung, B.~Giraud, P.~Vivet,
  A.~Levisse, and N.~Wehn, ``{RRAMSpec: A Design Space Exploration Framework
  for High Density Resistive RAM},'' in \emph{Embedded Computer Systems:
  Architectures, Modeling, and Simulation}, D.~N. Pnevmatikatos, M.~Pelcat, and
  M.~Jung, Eds.\hskip 1em plus 0.5em minus 0.4em\relax Cham: Springer
  International Publishing, 2019, pp. 34--47.

\bibitem{Lammie2020}
\BIBentryALTinterwordspacing
C.~Lammie, W.~Xiang, B.~Linares{-}Barranco, and M.~R. Azghadi, ``{MemTorch: An
  Open-source Simulation Framework for Memristive Deep Learning Systems},''
  \emph{CoRR}, vol. abs/2004.10971, 2020. [Online]. Available:
  \url{https://arxiv.org/abs/2004.10971}
\BIBentrySTDinterwordspacing

\bibitem{Rasch2021}
M.~J. Rasch, D.~Moreda, T.~Gokmen, M.~Le~Gallo, F.~Carta, C.~Goldberg,
  K.~El~Maghraoui, A.~Sebastian, and V.~Narayanan, ``{A Flexible and Fast
  PyTorch Toolkit for Simulating Training and Inference on Analog Crossbar
  Arrays},'' in \emph{Proceedings of the IEEE International Conference on
  Artificial Intelligence Circuits and Systems (AICAS)}, 2021.

\bibitem{Peng2021}
X.~Peng, S.~Huang, H.~Jiang, A.~Lu, and S.~Yu, ``{DNN+NeuroSim V2.0: An
  End-to-End Benchmarking Framework for Compute-in-Memory Accelerators for
  On-Chip Training},'' \emph{IEEE Transactions on Computer-Aided Design of
  Integrated Circuits and Systems}, vol.~40, no.~11, pp. 2306--2319, 2021.

\bibitem{Lammie2022}
C.~Lammie, W.~Xiang, and M.~{Rahimi Azghadi}, ``Modeling and simulating
  in-memory memristive deep learning systems: An overview of current efforts,''
  \emph{Array}, vol.~13, p. 100116, 2022.

\bibitem{Gale2019}
\BIBentryALTinterwordspacing
T.~Gale, E.~Elsen, and S.~Hooker, ``{The State of Sparsity in Deep Neural
  Networks},'' \emph{CoRR}, vol. abs/1902.09574, 2019. [Online]. Available:
  \url{http://arxiv.org/abs/1902.09574}
\BIBentrySTDinterwordspacing

\bibitem{eshraghian2021training}
J.~K. Eshraghian, M.~Ward, E.~Neftci, X.~Wang, G.~Lenz, G.~Dwivedi,
  M.~Bennamoun, D.~S. Jeong, and W.~D. Lu, ``{Training Spiking Neural Networks
  Using Lessons From Deep Learning},'' \emph{arXiv preprint arXiv:2109.12894},
  2021.

\bibitem{Pappalardo2021}
\BIBentryALTinterwordspacing
A.~Pappalardo, ``{Xilinx/Brevitas},'' 2021. [Online]. Available:
  \url{https://doi.org/10.5281/zenodo.3333552}
\BIBentrySTDinterwordspacing

\bibitem{Paszke2019}
A.~Paszke, S.~Gross, F.~Massa, A.~Lerer, J.~Bradbury, G.~Chanan, T.~Killeen,
  Z.~Lin, N.~Gimelshein, L.~Antiga, A.~Desmaison, A.~Kopf, E.~Yang, Z.~DeVito,
  M.~Raison, A.~Tejani, S.~Chilamkurthy, B.~Steiner, L.~Fang, J.~Bai, and
  S.~Chintala, ``{PyTorch: An Imperative Style, High-Performance Deep Learning
  Library},'' in \emph{Advances in Neural Information Processing Systems
  32}.\hskip 1em plus 0.5em minus 0.4em\relax Curran Associates, Inc., 2019,
  pp. 8024--8035.

\bibitem{Lammie2021}
C.~Lammie, M.~R. Azghadi, and D.~Ielmini, ``{Empirical Metal-Oxide RRAM Device
  Endurance and Retention Model for Deep Learning Simulations},''
  \emph{Semiconductor Science and Technology}, vol.~36, no.~6, p. 065003, 2021.

\bibitem{Simonyan2015}
K.~Simonyan and A.~Zisserman, ``{Very Deep Convolutional Networks for
  Large-Scale Image Recognition},'' in \emph{Proceedings of the International
  Conference on Learning Representations}, 2015.

\bibitem{Xie2017}
S.~Xie, R.~Girshick, P.~Dollár, Z.~Tu, and K.~He, ``{Aggregated Residual
  Transformations for Deep Neural Networks},'' in \emph{Proceedings of the IEEE
  Conference on Computer Vision and Pattern Recognition (CVPR)}, 2017, pp.
  5987--5995.

\bibitem{Sandler2018}
M.~Sandler, A.~Howard, M.~Zhu, A.~Zhmoginov, and L.-C. Chen, ``{MobileNetV2:
  Inverted Residuals and Linear Bottlenecks},'' in \emph{Proceedings of the
  IEEE/CVF Conference on Computer Vision and Pattern Recognition}, 2018, pp.
  4510--4520.

\bibitem{Szegedy2015}
C.~Szegedy, W.~Liu, Y.~Jia, P.~Sermanet, S.~Reed, D.~Anguelov, D.~Erhan,
  V.~Vanhoucke, and A.~Rabinovich, ``{Going Deeper With Convolutions},'' in
  \emph{Proceedings of the IEEE Conference on Computer Vision and Pattern
  Recognition (CVPR)}, 2015.

\bibitem{Shim2020}
W.~Shim, Y.~Luo, J.-s. Seo, and S.~Yu, ``{Impact of Read Disturb on Multilevel
  RRAM based Inference Engine: Experiments and Model Prediction},'' in
  \emph{Proceedings of the IEEE International Reliability Physics Symposium
  (IRPS)}, 2020.

\end{thebibliography}

\end{document}